\documentclass[aps,prd,reprint,groupedaddress]{revtex4-2}
\input epsf.sty
\usepackage{graphicx,amsmath,amssymb}

\def\lromn#1{\uppercase\expandafter{\romannumeral#1}}

\begin{document}


\title{
Stronger gravity in the early universe
}



\author{M. Yoshimura}
\affiliation{Research Institute for Interdisciplinary Science,
Okayama University \\
Tsushima-naka 3-1-1 Kita-ku Okayama
700-8530 Japan}


\date{\today}

\begin{abstract}
Scalar-tensor  theories of gravity that embrace conformal coupling 
to the scalar curvature
are the focal point of cosmology on discussions of  inflation 
and late-time accelerating universe.
Although there exists a stringent nucleo-synthesis constraint
on conformal gravity,
one can formulate how to evade this difficulty  by
modifying the standard  particle theory action consistently with the principles
of gauge invariant quantum field theory.
It is shown that stronger gravity at
early epochs of cosmological evolution than previously
thought of  is inevitable in a class of conformal gravity models.
This enhances discovery potentials of primordial gravitational wave emission
and primordial black hole formation.
The strong gravity effect may be enormous if massive clumps are energetically
dominated by cold dark matter made of inflaton field, and created black holes may
become a major candidate of cold dark matter. 

\end{abstract}


\maketitle



{\bf Introduction} \hspace{0.3cm}
It is natural to expect that both inflation at earliest epoch
of cosmological evolution and dark energy of
late-time acceleration  \cite{dark energy} have a common origin in a scalar degree of
freedom called inflaton \cite{cosmology}.
Since a temporarily stationary inflaton field needed in both cases
resembles a cosmological constant,
it would be ideal that the problem of fine-tuned cosmological constant  
\cite{lambda-problem}
is solved together with inflation and dark energy.

Scalar-tensor theory of gravity has
been a focal point in the advent of small, but finite dark energy.
Among them a class of conformal gravity theories
have appealing features, and many popular cosmology models
\cite{k-essence}, \cite{higgs-inflation}, \cite{linde et al}
of inflation and late-time acceleration  are interpreted 
as this class of theories.

In a recent publication \cite{cc relaxation} it was shown that
a special class of conformal gravity models has
a potential of solving the long-standing fine-tuning problem
of cosmological constant.
It was also pointed out that in these models
the cosmological mass variation of standard model particles occurs.
Historically, the problem of varying mass in cosmology is traced back to
the Jordan-Brans-Dicke (JBD) theory \cite{jbd}.
A partial motivation of  constructing this class of theory is 
the Dirac's large number hypothesis \cite{dirac large N}
proposed to explain the unnaturally large mass ratio
found in nature.
When a cosmological constant is added to the JBD theory,
the mass variation becomes inevitable \cite{varying mass}.
The Dirac's original idea is untenable due to the nucleo-synthesis constraint
explained below,
but the question remains on variation of the proton to the Planck mass ratio
$\approx 10^{-18}$.

The crucial question to be addressed to is how one can regulate
the pattern of mass variation in cosmology.
There exists a constraint on the mass ratio of proton to weak
W-boson from nucleo-synthesis
\cite{nucleo-synthesis bound}, \cite{gr tests},
which should be respected.
We shall formulate in the present work the method of modification in accordance with
the general principles of gauge invariant quantum field theory,
a basis of standard particle theory.

We use the natural unit of $\hbar = c = 1$ and the Boltzmann constant
$k_B= 1$ throughout the present work  unless otherwise stated.
The flat Friedman-Robertson-Walker (FRW) metric we use
is $ds^2 = dt^2 -a^2(t) (d\vec{x})^2$ with $a(t)$ the cosmic scale factor.

\vspace{0.5cm}
{\bf A class of conformal gravity under consideration} \hspace{0.3cm}
We start from formulation of  local quantum field theory in four space-time dimensions.
The lagrangian density we consider is given in what is called the Jordan metric frame \cite{jbd},
\begin{eqnarray}
&&
{\cal L} = {\cal L}_{g \chi} + {\cal L}_{\rm SM} 
\,, \hspace{0.5cm}
{\cal L}_{\rm SM} =  \sqrt{-g} L_{\rm SM}(\psi, g_{\mu \nu})
\,,
\\ &&
\hspace*{-0.3cm}
\frac{{\cal L}_{g \chi}}{ \sqrt{-g}} =  \left(  - M_{\rm P}^2 F(\chi) R(g_{\mu \nu})
- 2 M_{\rm P}^2\, \Lambda
+ \frac{1}{2} (\partial \chi)^2 - V(\chi)
\right)
\,, 
\nonumber \\ &&
\label {conformal gravity lagrangian}
\\ &&
M_{\rm P}^2 = \frac{1}{16 \pi G_N} \sim (1.72 \times 10^{18}\, {\rm GeV})^2
\,, \hspace{0.5cm}
\Lambda > 0
\,.
\end{eqnarray}
The spinless field $\chi$ plays the role of inflaton that mediates inflation
and late-time acceleration, while 
$L_{\rm SM}(\psi, g_{\mu \nu}) $ is the standard model lagrangian density, $\psi$
generically representing fields of standard particle theory; 
gauge bosons, Higgs boson, and fermions (leptons and quarks).
The conformal function $F(\chi)$ is assumed to be positive definite.
The cosmological constant $\Lambda$ is assumed to be positive.
The Jordan metric frame is suitable when four dimensional theory
descends from higher dimensional theories such as
Kaluza-Klein unification \cite{kk unification}
\cite{gh unification}, and superstring theories.

There are various possibilities for the potential function $V(\chi)$.
Many popular models \cite{k-essence}, \cite{higgs-inflation}, \cite{linde et al}
of non-minimal  kinetic
terms $K(\chi) (\partial \chi)^2 $ with a non-trivial function
$K$ in front  are transformed to conformal gravity of the type 
(\ref{conformal gravity lagrangian})
by a metric rescaling called Weyl transformation later elaborated.
Many remarks in the present work are applied to these models, as well.

Physically more transparent is the Einstein metric frame
obtained by a Weyl rescaling of the metric tensor,
$\bar{g}_{\mu \nu} = F(\chi) g_{\mu \nu}$ that eliminates $F-$factor
in front of the Ricci curvature in the Jordan frame.
To simplify our notation, we replace  the transformed metric $\bar{g}_{\mu \nu}  $ 
in the Einstein frame by $g_{\mu\nu}$, to derive
the scalar-tensor gravity part;
\begin{eqnarray}
&&
\frac{{\cal L}_{g\chi}}{ \sqrt{-g}} =
 \left(  - M_{\rm P}^2 R(g_{\mu \nu})
\right.
\nonumber \\ &&
\left.
+ \frac{5}{2 F(\chi)} (\partial \chi)^2 - \frac{1}{F^2(\chi)} 
(2 M_{\rm P}^2\, \Lambda + V(\chi) \right)
\,.
\label {inflaton part in e-frame}
\end{eqnarray}
In this frame the gravitational constant or
the Planck energy scale $M_{\rm P}$ is kept as an invariant constant
with cosmological evolution.
Different powers, $1/F$ and $1/F^2$, that appear in kinetic and potential terms
are determined by the presence or the absence of inverse metric tensor $g^{\mu\nu}$
in the Jordan frame.
The original cosmological constant in the Jordan frame becomes a
variable function $\Lambda/F(\chi)$ that is allowed to cosmologically evolve with
inflaton field $\chi$.

\vspace{0.5cm}
 {\bf Time evolution of inflaton field and Higgs boson mass}
\hspace{0.3cm}
The variational principle gives the inflaton field equation under the background of
the Friedman-Robertson-Walker metric,  
\begin{eqnarray}
&&
\ddot{\chi} +  3 \frac{\dot{a}}{a} \dot{\chi}
-  \frac{\partial_{\chi} F}{F } \dot{\chi}^2 
= - \partial_{\chi} V_{\rm eff}^{(E)}(\chi)
\,,
\\ &&
 V_{\rm eff}^{(E)}(\chi)
 = \frac{1}{5} \int_{\chi_*}^{\chi}  d\chi F(\chi) \partial_{\chi} V_{\chi g}^{(E)} (\chi)
\,,
\\ &&
 F \, \frac{\partial }{\partial \chi} V_{\chi g}^{(E)} (\chi)
=
\frac{1}{ F}
\left( \partial_{\chi}V - 2 (V + 2 M_{\rm P}^2 \Lambda) \frac{\partial_{\chi} F}{F}  \right)
\,.
\end{eqnarray}
The dot here means time derivative.
We considered spatially homogeneous mode, hence time-independent equation.
The effective potential $V_{\chi g}^{(E)} (\chi) $ 
was derived from in the force term of $\ddot{\chi}$
equation, and the original potential $V$ may include the centrifugal repulsive potential
when the multiple scalar field $\chi$ exhibits spontaneous symmetry breaking
as discussed in \cite{cc relaxation}.

A choice of quartic polynomials for two functions, $F(\chi)\,, V(\chi)$,
resolves in the simplest way the cosmological constant problem
along with realization of slow-roll inflation and late-time acceleration \cite{cc relaxation}.
We shall not restrict $F(\chi)\,, V(\chi)$ to these quartic forms
in the present work in order to treat a wider class of models.
But in this section we consider quartic polynomials as an example.
The leading behavior of potential at large field values is of a logarithmic type,
\begin{eqnarray}
&&
V_{\rm eff}^{(E)}(\chi) \approx \frac{M_{\rm P}^4}{5}
\left( {\rm constant}  - \frac{g}{\xi_4} \ln \frac{\chi}{\chi_*}
\right)
\,,
\label {effective potential at infty}
\end{eqnarray}
with $\chi_*$ a value of order the Planck energy $M_{\rm P}$
and $g\,, \xi_4$ coupling constants that appear in the potential $V$ and
the conformal function $F= \xi_4 (\chi/M_{\rm P})^4 + \cdots$.
After inflation a spontaneously broken symmetry is restored and 
the field $\chi$ rolls down very slowly to the field infinity, giving
a resolution of cosmological constant problem.
During this monotonic inflaton motion the effective inflaton mass $M_{\chi}$
changes from of order the Plank energy $O(10^{27})$ eV to of order the present Hubble 
constant $O(10^{-33})$ eV.
The product $M_{\chi} \chi$ is kept constant in the radiation-dominated epoch,
with a value $M_{\rm P} H_0 \sim ({\rm a\; few\; meV})^2$,
to give the present dark energy density of order $ ({\rm a\; few\; meV})^4$
as observed.
Derivation of these and other results for quartic polynomial model
is given in \cite{cc relaxation}.

If a conformal gravity of this sort is to describe a hot big-bang after
inflation, the inflaton field must couple to standard model particles
such as Higges boson, which further produces gauge bosons and light
fermions realizing a thermal equilibrium characterized by a single temperature.
If one takes the standard form of $L_{\rm SM}$,
one is led in the Einstein frame to different powers, 
 $1/F$ and $1/F^2$ for kinetic and potential terms, respectively,
exactly in the same way as in eq.(\ref{inflaton part in e-frame}) for the inflaton.
Inflaton coupling to Higgs doublet is given by the lagrangian density,
\begin{eqnarray}
&&
V_H^{(E)} =\frac{\lambda_H}{4} \,  F^{- 2} (| H|^2 - v^2 )^2
\,.
\end{eqnarray}
This gives both varying Higgs boson mass 
and Higgs coupling to inflaton.
The Higgs mass is modified from the standard result in general relativity 
$ \sqrt{2 \lambda_H} v$ to
$\sqrt{2 \lambda_H/\xi_4} \, (M_{\rm P}/\chi)^2\, v$.
Three-point vertex $\propto \chi H_0 H_0$ is important to realize
thermalized hot big-bang after inflation.

\vspace{0.5cm}
{\bf  Dispersion and the Einstein relation for freely moving particles}
\hspace{0.3cm}
Before we proceed, let us go back and find out the core part of the problem.
One may decompose the free field equation  into Fourier modes of the form,
$\propto e^{- i \omega t + \vec{q}\cdot \vec{x}}$, to derive a particle field $p$
under the FRW metric,
\begin{eqnarray}
&&
\omega^2 + i \,3 \frac{\dot{a}}{a} \omega -  i \, \frac{\dot{F}}{F} \omega
- \frac{q^2}{a^2} - \frac{m^2}{F^p} = 0
\,.
\label {dispersion relation}
\end{eqnarray}
Dot in this paper means the time derivative $d/dt$.
Relative weight difference in kinetic and mass terms is reflected
in the  $F-$factor power, $1/F^p$, of squared mass term. 
This is a dispersion relation between complex frequency (or energy) $\omega$ and
real 3-momentum $\vec{q}$.
During time span much shorter than variation time scales of $F$ and $a$,
 the analysis is simplest:
the imaginary part signifies decaying or frictional behavior, and let us
for the moment ignore this decay part, to derive a
dispersion relation among real parts of the right-hand side in (\ref{dispersion relation}),
\begin{eqnarray}
&&
\Re \omega_q = \sqrt{k^2+ \frac{m^2}{F^p} }
\,, \hspace{0.3cm}
\vec{k } = \frac{\vec{q}}{a}
\,.
\end{eqnarray}
The negative energy solution exists, as well.
The presence of cosmic scale factor $q^2/a^2$ in this relation
is well understood as the redshift of length scale associated with
cosmic expansion, and indeed by re-defining a physical momentum 
by $k = q/a$ we  recover the usual dispersion relation.

Both electroweak gauge bosons and Higgs boson (as shown above) have the power $p=1$.
For fermions one can square the Dirac spinor equation and
derive Klein-Gordon equation.
The Dirac mass follows the rule $\propto 1/F^2$, unlike W-boson
mass $\propto 1/F$.
Thus, its mass ratio is proportional to $1/F$.
Nucleo-synthesis is sensitive to a combination of
weak interaction parameters, $\nu =G_F m_p^2 \propto g_W^2 m_p^2/m_W^2$,
and concordance of theoretical calculation with observations
 places a bound on its variation, $|\delta \nu/\nu| < 0.06$,
\cite{nucleo-synthesis bound}, \cite{gr tests}.
The naive Weyl scaling from the Jordan frame
is thus ruled out from concordance of nucleo-synthesis
calculation with observations.
A special care on the metric frame in which the standard model is introduced
 must be taken in many popular models such as 
\cite{k-essence}, \cite{higgs-inflation}, \cite{linde et al}.
Otherwise, a thermalized hot universe after inflation might never occur in these models.

On the other hand, the stringent Oklo bound \cite{alpha variation}
is sensitive to the fine structure constant variation,
$|\delta \alpha/\alpha|) \leq O(10^{-7})$.
This bound is not applicable to models we discuss in the present work.

\vspace{0.5cm}
{\bf Theoretical framework for modifying the standard model lagrangian} \hspace{0.3cm}
We need firm theoretical principles to formulate
modification of the standard model lagrangian in the Jordan frame.
 Two principles to be respected are (1) the canonical quantization rule,
and (2) the gauge invariance.

Before spelling out these principles, it is useful to introduce
a fictitious counting rule applied to coordinate, its operation and field operators.
We assign counting numbers, $Q_F$, to these, 
as if they were conserved quantum numbers:
\begin{eqnarray}
&&
Q_F = -1; \hspace{0.2cm}
x_{\mu} \,, \; \partial_{\mu} = \frac{\partial}{\partial x^{\mu}}
\,, \; A_{\mu}
\,,
\\ &&
Q_F = 2 ; \hspace{0.2cm}
g^{\mu \nu} 
\,, \hspace{0.5cm}
Q_F = -2 ; \hspace{0.2cm}
g_{\mu \nu} 
\,,
\\ &&
Q_F = 0 ; \hspace{0.2cm}
H  \,, \; \chi  
\,, \; 
\psi_f
\,.
\end{eqnarray}
The rule is constructed such that 
\begin{eqnarray*}
&&
ds^2 = g_{\mu\nu} dx^{\mu}dx^{\nu} \,, 
\hspace{0.5cm}
g^{\mu\nu} \partial_{\mu}\chi \partial_{\nu}\chi
\,,
\end{eqnarray*}
are invariant with their $Q_F = 0$ among other working hypothesis
such as a well-defined $Q_F = -1$ for covariant derivative $\nabla_{\mu}$.
There is some ambiguity of $Q_F$ assignment on fermion field $\psi_f $, 
but this ambiguity is irrelevant in subsequent discussion.

One can form bosons from fermions as bound fermion-antifermion
pairs, hence it is imperative
to insert the same $F-$factor for boson commutators and fermion anti-commutators.
Hence the equal-time quantization rules for canonical conjugate
operators should satisfy a common
relation for spinless boson $\phi(x)$, any fermion $\psi(x)$,
and vector boson  fields $A^a_i\,, i =1,2,3$:
\begin{eqnarray}
&&
[ \phi(\vec{x}, t)\,, \nabla_0 \phi^{\dagger}(\vec{y}, t) ] =
 F^{\beta} \delta^{3} ( \vec{x} - \vec{y})
\,,
\label {quantization rule 1}
\\ &&
\{ \psi_s(\vec{x}, t) \,, \psi_t^{\dagger} (\vec{y}, t) \} = 
\delta_{st} \, F^{\beta} \delta^{3} ( \vec{x} - \vec{y})
\,,
\label {quantization rule 2}
\\ &&
[ A^a_i (\vec{x}, t)\,, \nabla_0 A^b_j(\vec{y}, t) ] =
\delta_{ab} \,g_{ij}  F^{\beta} \delta^{3} ( \vec{x} - \vec{y})
\,,
\label {quantization rule 3}
\end{eqnarray}
with $\nabla_{\mu}$ covariant derivative.
The canonical choice $\beta = 0$ is not excluded.
The metric tensor component
$g_{ij}$ should satisfy the relation $\partial^i g_{ij} = 0$,
since we quantize gauge fields in the Coulomb gauge,
but we shall keep it this way for simplicity.
The common power $\beta$ here is reflected
in creation and annihilation operators when the hamiltonian
is written as bilinear forms for the free field parts.

We now address the question of gauge invariance.
All masses in the standard model are generated by the Higgs mechanism
formulated in a gauge invariant manner.
There are five gauge invariant operators that appear in
the standard model lagrangian, denoted by \lromn1 $\sim$ \lromn5
below:
\begin{eqnarray}
&&
{\rm \lromn1}; \hspace{0.3cm}
L_{A} =
- \frac{1}{4}
g^{\mu \alpha } g^{\nu \beta } {\rm tr}\,
( \partial_{\mu} {\cal A}_{\nu} -  \partial_{\nu} {\cal A}_{\mu} 
+ i \frac{g}{2} [{\cal A}_{\mu}\,, {\cal A}_{\nu}])
\nonumber \\ &&
\times ( \partial_{\alpha} {\cal A}_{\beta} -  \partial_{\beta} {\cal A}_{\alpha} 
+ i \frac{g}{2} [{\cal A}_{\alpha}\,, {\cal A}_{\beta}]))
\end{eqnarray}
for non-Abelian and Abelian squared gauge field strength written
in the matrix form,
and for the Higgs kinetic term
\begin{eqnarray}
&&
{\rm \lromn2}; \hspace{0.3cm}
L_{dH} =
g^{\mu\nu}
\left[
\left(\partial_{\nu} - i( \frac{g' }{2} B_{\nu} + \frac{g}{2}\, \vec{\tau}\cdot\vec{A}_{\nu}) 
\right) \vec{H} \right]^{\dagger} 
\nonumber \\ &&
\times
\left(\partial_{\mu} - i( \frac{g' }{2}  B_{\mu} + \frac{g}{2}\, \vec{\tau}\cdot\vec{A}_{\mu}) 
\right) \vec{H}  
\,.
\end{eqnarray}
Higgs potential term is
\begin{eqnarray}
&&
{\rm \lromn3}; \hspace{0.3cm}
L_{H} =
- \frac{\lambda}{4} (|\vec{H}|^2 - v^2)^2
\,.
\end{eqnarray}
The fermion kinetic term is
\begin{eqnarray}
&&
{\rm \lromn4}; \hspace{0.3cm}
L_{df} =
\bar{\psi_f} i \Gamma_{\mu} \nabla^{\mu} \psi_f 
\,, 
\\ &&
\nabla^{\mu} = g^{\mu \nu} (\partial_{\nu} - i ( \frac{g'}{2} B_{\nu} 
+ \frac{g}{2} \vec{\tau}\cdot\vec{A}_{\nu}) 
\,.
\end{eqnarray}
Finally, fermion masses are generated from gauge invariant Yukawa coupling of Higgs field to fermion $\psi_f$:
\begin{eqnarray}
&&
{\rm \lromn5}; \hspace{0.3cm}
L_{f} =
y_f \left( \overline{\vec{\psi}_f}  \frac{1+ \gamma_5}{2} \psi_f\cdot \vec{H} + ({\rm h.c.})
\right)
\,.
\end{eqnarray}
All gauge invariant operators are arranged to have
a common $Q_F = 0$ charge;
namely they are $Q_F$ singlets.

Gauge invariance and canonical quantization rule of quantum field theory do not preclude
introducing arbitrary powers of $F$ in front of each gauge invariant operator.
Anticipating the Weyl rescaling, we count the number of 
the inverse metric $g^{\mu\nu}$ and multiply the
same number of $F-$factor to  gauge invariant operators that belong to
the same number of inverse metric.
Our general proposal  to modify the standard model lagrangian in the Jordan frame
is to use for $L_{\rm SM}/\sqrt{-g} $
\begin{eqnarray}
&&
F^{p_1}  L_A + F^{p_2+1} (L_{dH} + L_{df}) + F^{p_3+2} ( L_{H} + L_f)
\,.
\end{eqnarray}
The Weyl rescaling to the Einstein metric frame changes these powers
to $p_1 \,, p_2 \,, p_3 $.
For particle energies and masses the factor $\beta$ arising from proper normalization
of creation and annihilation operators 
from (\ref{quantization rule 1}) $\sim $ (\ref{quantization rule 3})
should also be taken into account.
Nucleo-synthesis  places a constraint:  $ (\beta + p_2)/2 = \beta + p_3$.
A parametrization that takes this into account is
$ (\beta + p_1)/2 = - \epsilon_1\,, (\beta + p_2)/2 = \beta + p_3 = - \epsilon_2 $.

In the following table we list $F-$powers of
 particle energies, masses and the energy to mass ratio
in the Einstein frame,
when the hamiltonian is written as bilinear forms of creation and annihilation
operators.

\vspace{0.5cm}
\begin{tabular}{cccc} \hline 
particles & energy power & mass power & power of $E/m$ \\ \hline
$W\,, Z$  & $F^{ -\epsilon_1} \, (F^{-1} ) $ & 
$F^{ -\epsilon_2} \, (F^{-1} )$   & $F^{ \epsilon_2- \epsilon_1} \, (1)$ \\ 
$H_0 $ & $F^{-\epsilon_2} \,(F^{-1} ) )$  
& $F^{-\epsilon_2} \, (F^{ -1} )$ & $1 \, (1)$ \\ 
fermions & $F^{ -\epsilon_2 } \, (F^{-2} )$ &  
$F^{-\epsilon_2 } \, (F^{-2} )$ &  $1\, (1)$ \\ 
inflaton & $F^{-1/2} $  & $F^{-1/2 } $ & $1 $ \\  \hline
\end{tabular}
\vspace{0.5cm}

For reference we also listed in the parenthesis powers given by
the naive Weyl rescaling, which however contradicts nucleo-synthesis constraint.

A further restriction to maintain the common Einstein relation $E= m c^2$ 
(c being the light velocity) to all particles is desirable, and it
leads to $\epsilon_1 = \epsilon_2 = \epsilon$.
The Jordan-frame lagrangian is then
\begin{eqnarray}
&&
\hspace*{-0.5cm}
\frac{{\cal L}_{\rm SM}^{(J)} } {\sqrt{-g } } = F^{-2\epsilon } \left(
F^{2}  L_A + F (L_{dH} + L_{df}) +  L_{H} + L_f \right)
\,,
\label {jordan frame modification}
\end{eqnarray}
taking into account the change of quantization rule.
When $\epsilon=0$, the Einstein frame lagrangian density is $\sqrt{-g}$
times the standard model quantity,
$ L_A + L_{dH} + L_{df} +  L_{H} + L_f $.
In this case particle energies and masses are exactly the same
as those of GR, except the inflaton.
But this solution is not favored, because it is impossible to
solve the cosmological constant problem in
the standard model lagrantian.
A finite positive or negative $\epsilon$ value, however small it is, 
solves the  dynamical cosmological constant towards zero
in a class of conformal gravity models \cite{cc relaxation}.

\vspace{0.5cm}
 {\bf Stronger gravity at early epochs}
\hspace{0.3cm}
We formulated the problem of the dispersion relation
and the mass variation in the Einstein metric frame in which the
gravitational constant is kept invariant with the cosmic evolution.
The dimensionless measure of gravity given by $G_N M^2$ or $G_N E^2$
for a  body of mass $M$ and energy $E$ ($G_N E^2$ being relevant when the
major part of constituent particles move with relativistic
velocities) changes with the cosmological evolution.
The gravity strength thus defined is different,  depending on whether the
major component of the clump mass arises from
ordinary baryons or  from CDM (cold dark matter) inflatons.
As in the model of \cite{cc relaxation},
we assume for definiteness that the inflaton increases monotonically towards
the field infinity, its dynamical mass  following $\propto \chi^2 \propto (z+1)^{4}$.

First, when the clump mass is energetically dominated by CDM inflaton,
\begin{eqnarray}
&&
G_N M^2\,, \; G_N E^2 \propto F^{-2 }(\chi) \propto   (z+1)^{ 8}
\,,
\end{eqnarray}
Irrespective of the sign and magnitude of $ \epsilon$, the growth rate of the effective
gravitational strength towards earlier epochs is enormous for CDM clumps.

We could think of three major epochs where strong gravity manifests itself:
(1) stochastic gravitational wave (GW) backgrounds
throughout the entire cosmological history \cite{stochastic gw backgr}, 
(2) the preheating stage of inflationary epoch, \cite{gw preheating},
(3) the first order electroweak phase transition
that may take place in extended Higgs models \cite{1st order pt}.
In (2) and (3) both GW emission and primordial black hole 
formation \cite{pbh}  may be copious.
It is likely that even a tiny gaussian tail of fluctuation
$M_{\rm CDM} (0)$ provides a very strong gravity to CDM clumps.
In the model of \cite{cc relaxation} the total amount of
primordial black holes is limited by $O({\rm  meV})^4$,
hence created black holes do not over-close the universe.
We feel it pressing to more seriously study gravitational collapse of 
the inflaton cold dark matter in the radiation-dominated universe.

Even at modest redshifts a high statistic data of
GW emission from black hole mergers may provide a
crucial test of strong gravity effect discussed here.
In particular, the frequency distribution of events in terms of
merger masses and redshifts have characteristic features of
dependence, $\propto M_1 M_2$ and $\propto (z+1)^8$.
One has to distinguish the primordial origin from
astrophysical origins in data analysis.

Gravity strength of clump mass $M$  made of baryonic  matter
 changes according to
\begin{eqnarray}
&&
G_N M^2 \propto F^{- 4\epsilon}(\chi)  \propto   (z+1)^{16 \epsilon}
\,.
\end{eqnarray}
Only when $ \epsilon$ is positive and not too small,
one can expect a substantial growth of gravitational strength
at early epochs.

We note a promising opportunity in the near future.
High statistics observations of
GW emission from  neutron star mergers \cite{gw from neutron star merger}
at different redshifts
may provide  an observational hint on the important parameter $ \epsilon$,
noting the 
gravity strength $G_N M_{NS}^2 (z+1)^{16 \epsilon } \,, M_{NS} \sim 1.3 \times$
the solar mass.

\vspace{0.5cm}
{\bf Summary and outlook} \hspace{0.3cm}
General relativity has been a remarkable success including recent
detections of gravitational wave emission from merging black hole binaries
\cite{gr tests from gw detection}.
Nevertheless, there is a deep conundrum related to
a small, but finite cosmological dark energy of
late-time evolution, which seems to require the scalar degree of freedom,
presumably also related to inflation.
Resolution of the fine-tuned cosmological constant problem
along with other cosmological conundrums
may find a solution in a class of conformal gravity theories,
as outlined in \cite{cc relaxation}.

We studied in the present work
consequences of cosmology models based on, or
interpreted as,  a class of conformal gravity theories.
Severe constraint from nucleo-synthesis restricts
how the inflaton field couples to standard model particles:
the mass ratio of proton to W-boson must not change with
cosmological evolution.
We formulated how to evade this difficulty by
modifying the standard model lagrangian in the Jordan frame and the canonical
quantization rule of quantized fields.
Our simplest solution is to introduce a slight modification of the ordinary standard
model lagrangian after the Weyl rescaling to the Einstein metric frame,
and not in the Jordan frame.
Many popular models interpreted as a sort of conformal gravity
must obey the same rule if they are to describe a hot big-bang after inflation.

A sacrifice to pay, or an exciting possibility to the future,  is 
the unexpected strong gravity in the early universe.
It predicts much stronger gravitational wave emission 
and much more copious primordial black hole formation
than GR predicts, if clumpy parts
of fluctuation are primarily made of inflaton cold dark matter.

\vspace{0.5cm}
\begin{acknowledgments}
This research was partially
 supported by Grant-in-Aid   21K03575   from the Japanese
 Ministry of Education, Culture, Sports, Science, and Technology.

\end{acknowledgments}

\end{document}